\documentclass[preprint,showpacs,amsmath,amssymb]{revtex4}
\usepackage{graphicx}%include figure files
\usepackage{dcolumn}%alingn table columns on decimal point
\usepackage{bm}%bold math
\begin{document}
\title{Band structure and magnetotransport of a two-dimensional electron gas 
in the presence of %Rashba and Dresselhaus 
spin-orbit interaction}
\author{X. F. Wang}
\email{xuefeng@alcor.concordia.ca}
\author{P. Vasilopoulos}
\email{takis@alcor.concordia.ca}
\affiliation{Department of Physics, Concordia University\\
1455 de Maisonneuve  Ouest, Montr\'{e}al, Qu\'{e}bec, Canada, H3G 1M8}
\begin{abstract}
The  band structure and magnetotransport of a two-dimensional electron gas (2DEG), in the
presence of the Rashba (RSOI) and Dresselhaus (DSOI) terms of the spin-orbit interaction
and of a perpendicular magnetic field, is investigated.
Exact and approximate analytical expressions for the band structure are obtained and used to
calculate the density of states (DOS) and the longitudinal magnetoresitivity  assuming a Gaussian
type of level broadening. The interplay between the Zeeman coupling and the two terms of the SOI
is discussed. If the strengths $\alpha$ and $ \beta$, of  the RSOI and DSOI, respectively,
are equal and the $g$ factor vanishes, the two spin states are degenerate  and a shifted
Landau-level structure appears. With the increase of the difference $\alpha- \beta$, a novel
beating pattern of the DOS and of  the Shubnikov-de Haas (SdH) oscillations appears distinctly
different from that occurring when one of these strengths vanishes.
\end{abstract}
\date{\today}
\maketitle
\section{introduction}
The spin-orbit interaction (SOI) in semiconductor nanostructures has attracted extensive
attention in the past years mostly due to
its potential applications in manipulating electron spins in electronic devices and
its ample physical characteristics hidden and unexplored previously. The SOI is a
relativistic coupling between the spin and momentum of an electron under an external electric
field. \cite{wink}  This external electric field can be the slope of the band structure or the
averaged crystal field due to an asymmetry of the bulk semiconductor crystal.
The former results in the Rashba SOI (RSOI) \cite{bych} and the later in the Dresselhaus SOI
(DSOI). \cite{dres} In semiconductor heterostructures an additional DSOI arises from
the asymmetry of the material growth at the interfaces that is different from
that due to the crystal asymmetry in bulk materials.
\cite{dyak,verv,ross} 

The RSOI is rotationally symmetric in a 2DEG and normally dominates in 2DEG's of many
narrow-gap semiconductors such as InGaAs/AlGaAs. \cite{wang2} This has
been confirmed in experiments \cite{luo,koga}. Recent experiments, however,
show that the DSOI makes an important contribution to the SOI and can be dominant
in 2DEG's of materials shuch as InSb/InAlSb. \cite{gani, here}
In a system without magnetic field, it has been shown that the DSOI introduces
a strong anisotropy to the electronic energy band and the electron transport. \cite{gani,schl1}
An interesting phenomenon in this case is that $k$-independent eigenspinors appear when the
strengths  $\alpha$ and $ \beta$, of  the RSOI and DSOI, respectively, are equal. \cite{schl1}

It is well known that under a perpendicular magnetic field the Hamiltonian
can be solved exactly if only one of the RSOI or the DSOI terms exists.
Two branches of energy levels with unequal separations between adjacent
levels develop and lead to a beating pattern of the SdH oscillations
observed in  moderate magnetic fields. \cite{luo,koga,nitt,wink,wang}
By measuring the position of the nodes of the beating pattern, one can
estimate the strength $\alpha$ or $ \beta$ of the SOI. The situation is much less
clear when both terms of the SOI are present as the number of pertinent studies
is limited \cite{gani}. In particular,
we are not aware of any treatment of the band structure and SdH oscillations of a 2DEG
under a perpendicular magnetic field when  both terms of the SOI  are present. We provide
one in  this paper and organize it as follows. In Sec. II we provide exact and approximate
results for the energy spectrum and the DOS. In Sec. III we present  analytical and
numerical results for the transport coefficients, and in Sec. V concluding remarks.

\section{Eigenvectors, eigenvalues, and density of states}

We consider a 2DEG in the $(x-y)$ plane and a magnetic field along the $z$
direction. In the Landau gauge $\vec{A}=(0,Bx,0)$ the one-electron
Hamiltonian $H=H_0+H_\alpha+H_\beta$ including
the SOI  Rashba term, $H_\alpha$,  and the  Dresselhaus SOI term, $H_\beta$,
reads \cite{bych,ross}
\begin{eqnarray}
H&=&\frac{(\hat{{\bf p}}+e{\bf A})^{2}}{2m^{\ast }} +
\frac{1}{2}g_s\mu_{B}B\sigma _{z},+H_\alpha+H_\beta \\ \nonumber
H_\alpha &=&\frac{\alpha}{\hbar}\sigma_x(\hat{p}_y+eA_y)
-\frac{\alpha }{\hbar}\sigma_y(\hat{p}_x+eA_x),\\ \nonumber
H_\beta &=&\frac{\beta}{\hbar}\sigma_x(\hat{p}_x+eA_x)
-\frac{\beta}{\hbar}\sigma_y(\hat{p}_y+eA_y),
\label{ham}
\end{eqnarray}
where $\hat{\bf p}=(\hat{p}_x,\hat{p}_y)$ is the momentum operator of the electrons,
$m^\ast$  their effective mass, $g_s$ the Zeeman factor,
$\mu _{B}$ the Bohr magneton, ${\bf \sigma}=(\sigma _{x},\sigma _{y},\sigma _{z})$
the Pauli spin matrix, and $\alpha$,  
$\beta$ the strengths of the Rashba and Dresselhaus terms, respectively.

In the absence of SOI the eigenstates of the Hamiltonian are the Landau states
$|n,\sigma \rangle$, $n=0,1,2,\cdots$, with energy
$\varepsilon_n^\sigma=(n+1/2)\hbar\omega_c+\sigma g_s\mu_{B}B/2$,
and wave function $\langle {\bf r}|n,\sigma \rangle
=e^{ik_{y}y}\phi _{n}(x+x_{c})|\sigma\rangle/\sqrt{L_y}$.
$L_y$ is the length of the system along the $y$ direction,
$\phi _{n}(x+x_{c})=e^{-(x+x_{c})^{2}/2l_{c}^{2}}
H_{n}( (x+x_{c})/l_{c})/\sqrt{\sqrt{\pi}2^{n}n!l_{c}}$ the  harmonic oscillator function,
$\omega _{c}=eB/m^{\ast}$ the cyclotron frequency, $l_{c} =(\hbar/m^{\ast}\omega _{c})^{1/2}$
the radius of the cyclotron orbit centered at $-x_{c}=-l_{c}^{2}k_{x}$, $n=0,1,2,\cdots$  the
Landau-level index, and $|\sigma\rangle=|\pm\rangle$ the electron spin written as the row
vector $|+\rangle = (0,1)$   if the spin points up and as $|+\rangle = (1,0)$
if it points down.

Generally, the SOI groups the Landau states into two groups $|n,\sigma_n\rangle$ and
$|n,-\sigma_n\rangle$ with $\sigma_n=(-1)^n$ and couples the states with each other
in each group. Assuming  eigenstates in the form of
$\Psi=\sum_n C_n^\sigma |n,\sigma\rangle$, the secular equation $H\Psi=E\Psi$ 
%%%
leads to the system of equations
\begin{eqnarray}
\sqrt{n}\epsilon_\alpha C_{n-1}^{+}+(\varepsilon_{n}^{-}-E)C_{n}^{-}-i
\sqrt{n+1}\epsilon_\beta C_{n+1}^{+} &=&0 \\ \nonumber
i\sqrt{n+1}\epsilon_\beta C_{n}^{-}+(\varepsilon_{n+1}^{+}-E)C_{n+1}^{+}+
\sqrt{n+2}\epsilon_\alpha C_{n+2}^{-} &=&0,
\end{eqnarray}
with $n$ odd for the group $|n,\sigma_n\rangle$ and even for the group $|n,-\sigma_n\rangle$. Here $\epsilon_\alpha=\sqrt{2}\alpha/l_c$ and $\epsilon_\beta=\sqrt{2}\beta/l_c$. This means that the Hamiltonian matrix in the Landau space reduces  to two
independent, infinitely dimensional matrices. 
The matrix corresponding to the group $|n,\sigma_n\rangle$ reads 
\begin{equation}
\left[ 
\begin{array}{ccccccc}
\varepsilon_{0}^{+} & \epsilon_\alpha & 0 & 0 & 0 & 0 & \cdots  \\ 
\epsilon_\alpha & \varepsilon_{1}^{-} & -i\sqrt{2}\epsilon_\beta & 0 & 0
& 0 & \cdots  \\ 
0 & i\sqrt{2}\epsilon_\beta & \varepsilon_{2}^{+} & \sqrt{3}\epsilon_\alpha & 0 & 0 & \cdots  \\ 
0 & 0 & \sqrt{3}\epsilon_\alpha & \varepsilon_{3}^{-} & -i2\epsilon_\beta & 0 & \cdots  \\ 
0 & 0 & 0 & i2\epsilon_\beta & \varepsilon_{4}^{+} & \sqrt{5}\epsilon_\alpha & \cdots  \\ 
\cdots  & \cdots  & \cdots  & \cdots  & \cdots  & \cdots  & \cdots 
\end{array}
\right] 
\label{matrixa}
\end{equation}
and the one corresponding to the group $|n,-\sigma_n\rangle$ 
\begin{equation}
\left[ 
\begin{array}{ccccccc}
\varepsilon_{0}^{-} & -i\epsilon_\beta & 0 & 0 & 0 & 0 & \cdots  \\ 
i\epsilon_\beta & \varepsilon_{1}^{+} & \sqrt{2}\epsilon_\alpha & 0 & 0 & 
0 & \cdots  \\ 
0 & \sqrt{2}\epsilon_\alpha & \varepsilon_{2}^{-} & -i\sqrt{3}\epsilon_\beta & 0 & 0 & \cdots  \\ 
0 & 0 & i\sqrt{3}\epsilon_\beta & \varepsilon_{3}^{+} & 2\epsilon_\alpha & 0 & \cdots  \\ 
0 & 0 & 0 & 2\epsilon_\alpha & \varepsilon_{4}^{-} & -i\sqrt{5}\epsilon_\beta & \cdots  \\ 
\cdots  & \cdots  & \cdots  & \cdots  & \cdots  & \cdots  & \cdots 
\end{array} 
\right]. 
\label{matrixb}
\end{equation}

\begin{figure}[tpb]
\vspace{-5cm}
\includegraphics*[width=120mm]{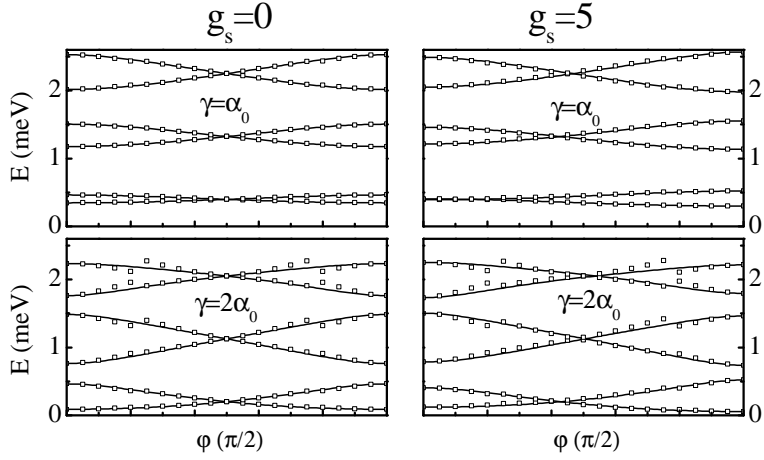}
\vspace{-5cm}
\caption{Energy spectrum of a 2DEG vs the angle $\varphi$  
($ \tan\varphi=\alpha/\beta$)  for two values of $\gamma=(\alpha^2+\beta^{2})^{1/2}$
in a perpendicular magnetic field $B=0.4$ T.
The solid curves denote the  numerical result,  obtained by diagonalizing the Hamiltonian matrices (\ref{matrixa}) and (\ref{matrixb}),  and the empty squares the approximate analytical
result described by Eq. (\ref{e2}) for $\tan\varphi < 1/3$, its $\beta\gg \alpha$ counterpart for
$\tan\varphi > 3$ , and by Eq. (\ref{e4}) for $1/3 < \tan\varphi < 3$.
On  the left panels  the Zeeman term is neglected, on the right ones it is not ($g_s=5$).
The effective electron mass used is $m^\ast=0.05 m_0$.} 
\label{fig1}
\end{figure}

The energy spectrum of Eqs. (\ref{matrixa}) and (\ref{matrixb}) can be obtained numerically
by truncating the matrix dimensions while including a sufficient number of Landau levels;  the resulting energy spectrum will be referred to as the exact one in this paper. For  $\alpha$
and $\beta$ varying independently, we study the spectrum in the  ($\alpha$-$\beta$)   plane and define an angle $\varphi$ at any point ($\alpha,\beta$)
such that $\alpha=\gamma\cos\varphi$ and $\beta=\gamma \sin\varphi$ with
$\gamma=(\alpha^2+\beta^{2})^{1/2}$. In Fig. \ref{fig1} we show the 
spectrum (solid curves) as a function of the angle $\varphi$, in units of $\pi/2$, at various
$\gamma$ values. 
%%%
The parameter $\alpha_0$ is equal to $10^{-11}$eVm.
We consider only positive $\alpha$ and $\beta$ so that $0 < \varphi < \pi/2$.  Results for negative $\alpha$ or $\beta$ can be obtained by symmetry.
The  results for $g_s=0$ and $g_s=5$ shown present the effect of the Zeeman term. The value of $g_s$ varies from sample to sample in the literature but we use $g_s=5$
since a recent experiment demonstrated a $g_s$ range from 2 to 5 in gate-controlled InGaAs/InAlAs quantum wells. \cite{nit2} The SOI splits the Landau levels in two branches, referred to as the $+$ and $-$ branches, which cross each other
when  $\varphi$ is varied. The two spin branches  in each Landau level are degenerate for
$\alpha=\beta$ or $\varphi=\pi/4$ if  the Zeeman term is absent; then  energy-branch crossing shifts with increasing $g_s$. This energy degeneracy can occur  for several values of $\varphi$ as shown by the upper curves in the plots with $\gamma=2\alpha_0$.

A better insight into the problem is obtained by an analytic energy spectrum. This
can facilitate the study of other properties, e.g., the transport properties of this system.
It is well known \cite{wang} that the Hamiltonian (\ref{ham}) has an exact solution if
$\alpha$ or $\beta$ vanishes. Here we will show
that an exact result is obtained also for $\alpha=\beta$, if we neglect the Zeeman term, and
 approximate perturbative results follow  for $\alpha/\beta \ll$ or $\gg 1$ and $\alpha\approx\beta$. 
 
%Fig.1

\subsection{Highly unequal strenths: $\alpha<<\beta$ or $\beta<<\alpha$}

For $\beta=0$ the Hamiltonian $H_0+H_\alpha$
in the Landau space,  cf. Eqs.({\ref{matrixa}) and (\ref{matrixb}),
reduces to $2\times 2$ matrices with exact
eigenvalues \cite{bych,wang,shen}
\begin{equation}
{\cal E}_n^\sigma=\hbar\omega_c(n+\sigma\sqrt{(1-g)^2+8n k^2_\alpha l^2_c}/2)
\label{e1}
\end{equation}
and  eigenfunction
\begin{equation}
\Psi^\sigma_n=\frac{e^{ik_y y}}{\sqrt{L_y}}
\left(
\begin{array}{c}
\phi_{n-1}(x+x_c)\sin\theta^\sigma_n \\
\phi_n(x+x_c)\cos\theta^\sigma_n \\
\end{array}
\right).
\label{w1}
\end{equation}
Here
$n=0,1,2,\cdots$ for $\sigma=1$ and $n=1,2,3,\cdots$ for $\sigma=-1$,
$g=g_sm^\ast/m_0$ with $m_0$ the free-electron mass,
$k_\alpha=\alpha m^\ast/\hbar^2$, $\theta^-_n=\theta^+_n-\pi/2$,
and
\begin{equation}
\tan\theta^+_n=\frac{2\sqrt{2n}k_\alpha l_{c}}{1-g+\sqrt{(1-g)^{2}
+8nk^2_\alpha l_{c}^{2}}}.
\end{equation}

For $\beta<<\alpha$, we can treat $H_\beta$ as a perturbation of $H_0+H_\alpha$.
Neglecting the cross terms $\propto\alpha\beta$, the matrix elements of $H_\beta$ are
\begin{equation}
\langle\Psi^\sigma_m|H_\beta |\Psi^{\sigma'}_n\rangle
=i(\sigma'-\sigma)\sqrt{(m+\sigma)}(\epsilon_\beta/2)
\delta_{m,n-2\sigma}. 
\end{equation}
Then the total Hamiltonian is a matrix composed of   diagonal elements 
%%%%?
$\varepsilon^-_1$
and a series of $2\times 2$ diagonal 
blocks. The  eigenvalues are
\begin{equation}
E_t^\sigma=({\cal E}^-_{t+1}+{\cal E}^+_{t-1})/2
+(\sigma/2)\sqrt{({\cal E}^-_{t+1}-{\cal E}^+_{t-1})^2+4t\epsilon_\beta^2};
\label{e2}
\end{equation}
the corresponding wave functions read
\begin{equation}
f^\sigma_t=-i\Psi^+_{t-1}\sin\gamma^\sigma_t +\Psi^-_{t+1}\cos\gamma^\sigma_{t},
\label{w2}
\end{equation}
with $\tan\gamma^\sigma_t=(2\sqrt{t}\epsilon_\beta)/[\sqrt{({\cal E}^-_{t+1}-{\cal E}^+_{t-1})^2
+4t\epsilon_\beta^2}+\sigma({\cal E}^-_{t+1}-{\cal E}^+_{t-1})]$ and $t=1,2,3,\cdots$ for $\sigma=1$ or $t=0,1,2,\cdots$ for $\sigma=-1$.

The  results corresponding to $\beta \gg \alpha$ can be obtained in
the same way. The  energy spectrum is
still given by Eqs.(\ref{e1}) and (\ref{e2}) with $\alpha$ and $\beta$ interchanged and $(1-g)$ replaced by $(1+g)$. The wave  functions are
given by Eq. (\ref{w1}), with $\sin\theta^\sigma_n$ replaced by $-i\sin\theta^\sigma_n$, 
and by Eq. (\ref{w2}) with $-i\sin\gamma^\sigma_n$ replaced by $\sin\gamma^\sigma_n$.
Equations (\ref{e2}) and (\ref{w2}) and their $\beta \gg \alpha$ counterpart will
be used as an analytical approximation for
$|\beta-\alpha|/(\alpha+\beta) > 0.5$ or $\tan\varphi > 3$ and $1/3 < \tan\varphi$.

\subsection{Equal ($\alpha=\beta$) or approximately equal ($\alpha\approx\beta$) strengths} 

For $\alpha=\beta$ the Hamiltonian $H=H_0+H_\alpha+H_\beta$ without the Zeeman term can be diagonalized
by a unitary transformation $U^\dag H U$ with 
\begin{equation}
U=\left[
\begin{array}{cc}
(1+i)/2 &-(1+i)/2 \\
 1/\sqrt{2}& 1/\sqrt{2}
\end{array}
\right].
\end{equation}
The nonzero elements of the diagonalized Hamiltonian read
\begin{equation}
H_{\sigma,\sigma}=(\hat{p}_x+\sigma\sqrt{2}\hbar k_s)^2/2m^\ast
+(\hat{p}_y+\sigma\sqrt{2}\hbar k_s+eBx)^2/2m^\ast-2\hbar^2 k_s^2/m^\ast
\label{diag}
\end{equation}
with $k_s=am^\ast/2\hbar^2$ and  $a=(\alpha+\beta)/2$.
This   Hamiltonian is very close to that of a displaced harmonic oscillator and we attempt
solutions in the form
$\Psi^\sigma(x,y)=e^{ik_y y}e^{-i\sigma\sqrt{2}k_s x}\phi(x)$ for the spin branch $\sigma$.
The wave function of this Hamiltonian,
in the Landau gauge, is then obtained as
\begin{equation}
\Psi^\sigma_n=\frac{e^{ik_y y}}{\sqrt{L_y}}
e^{-i\sigma\sqrt{2}k_s x}
\phi_n(x+x_c+\sigma x_s)
\left(
\begin{array}{c}
\sigma(1+i)/2 \\
1/\sqrt{2}
\end{array}
\right)
\label{w3}
\end{equation}
with $x_c=l^2_c k_y$,  $x_s=\sqrt{2}l^2_c k_s$ 
and eigenvalue
\begin{equation}
{\cal E}^\sigma_n=(n+1/2)\hbar\omega_c-2\hbar^2 k_s^2/m^\ast.
\label{e3}
\end{equation}

For $\alpha\neq\beta$  the Hamiltonian given by Eq. (\ref{ham}) after the unitary transformation has diagonal and nondiagonal elements. The diagonal ones are 
given by Eq. (\ref{diag}) and the nondiagonal ones by
\begin{equation}
H_{\sigma,-\sigma}=i\sigma\sqrt{2}b(\hat{p}_y/\hbar-\hat{p}_x/\hbar+x/l^2_c)-g_s\mu_BB/2
\end{equation}
with $b=(\alpha-\beta)/2$. For $\alpha\approx\beta$ and a Zeeman term  weak compared to the Landau energy,  which is usually  the case, we can treat the nondiagonal elements as a perturbation and neglect the coupling between levels of different index in Eq.   (\ref{w3}). The resulting approximate energy spectrum reads
\begin{equation}
E^\sigma_n=(n+1/2)\hbar\omega_c-2\hbar^2k^2_s/m^\ast
+\sigma\eta_n/2
\label{e4}
\end{equation}
with $\eta_n=2\sqrt{2}b\Delta_n-2\Sigma_n$ the spin splitting of the $n$-th level,
%%%
$\Delta_n=2Z_se^{-2Z_s^2}[L^{1}_n(4Z^2_s)
+L_{n-1}^{1}(4Z^2_s)]/l_c$, $\Sigma_n=g_s\mu_BBe^{-Z_s^2}L_n(2Z_s^2)$,
 $Z_s=\sqrt{2}k_sl_c$, and $L_n^m(x)$ the Laguerre polynomial. The corresponding approximate wave function is given by
\begin{equation}
f^\sigma_n=(\Psi_n^++\sigma\Psi_n^-)/\sqrt{2}.
\label{w4}
\end{equation}
Eq. (\ref{e4}) is used as the approximate analytical energy spectrum in the regime
$|\alpha-\beta|/(\alpha+\beta) < 0.5$ or $1/3 < \tan \varphi < 3$.

The approximate  energy spectrum given by Eqs. (\ref{e2}) and (\ref{e4})
is shown in Fig. \ref{fig1} by the empty squares.  It agrees well in the whole range of $\varphi$
with the exact result ( solid curves) when the overall SOI strength $\gamma$ is weak.
Since most of the existing measurements show SOI strengths in the order of $\alpha_0$,
our approximate energy spectrum is pertinent to many experiments.
To show more clearly the discrepancy between the 'exact' result and the approximate one,
we use a magnetic field $B=0.4$ T in Fig. \ref{fig1} which is twice  as strong as
that used in  later discussions.

%Fig.2
\begin{figure}[tpb]
\vspace{-1cm}
\includegraphics*[width=160mm, height=160mm]{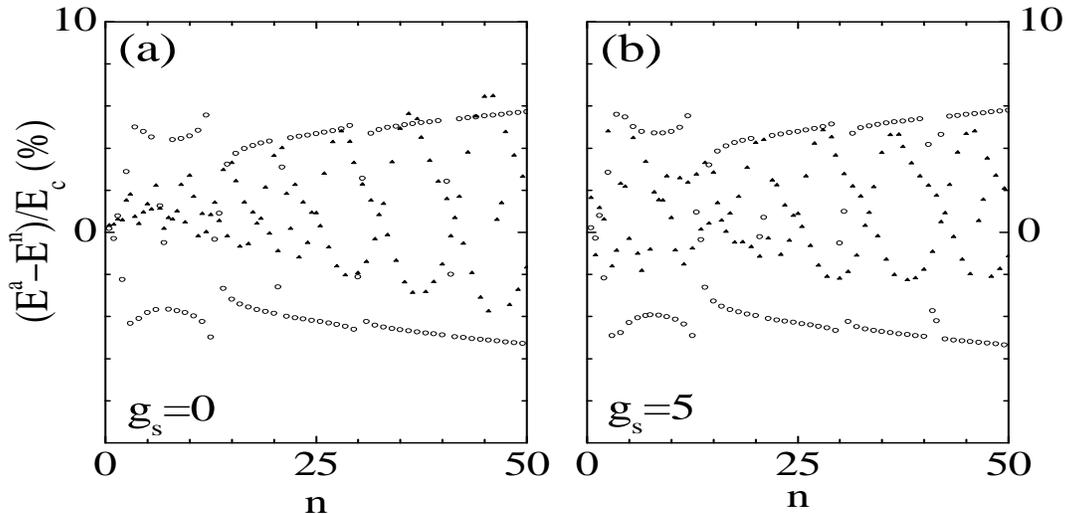}
\vspace{-8cm}
\caption{Energy difference between the approximate analytical $E^a$ and numerical $E^n$ results, in units of the Landau energy $E_c=\hbar\omega_c$,
vs the Landau-level index $n$. The triangles show results for $\alpha=2.2\alpha_0$ and $\beta=1.8\alpha_0$, and the empty circles for $\alpha=\alpha_0$ and $\beta=0.2\alpha_0$.} 
\label{fig2}
\end{figure}

To see how accurate  the approximate analytical result $E^a$ is for  higher-index subbands,
which may be occupied in weak magnetic fields, in Fig. \ref{fig2} we plot the energy difference $E^a-E^n$ between it and the exact numerical result $E^n$ normalized to the Landau energy $E_c=\hbar\omega_c$ for up to 100 levels or $n=50$ at a magnetic field $B=0.2$ T.
On the average the error introduced by using the approximate result increases with the index $n$ and the SOI strength.  In the absence of the Zeeman term, cf. Fig. \ref{fig2}(a), the approximate formula Eq. (\ref{e2}), corresponding to the empty circles, overestimates the subband
energies of one branch while underestimates those of the other and the errors are almost proportional to the subband index $n$ for big $n$. Referring to Fig. \ref{fig1},
we find that the energy gap between the two branches estimated by Eq. (\ref{e2}) is generally narrower than it should be. The errors resulting from Eq. (\ref{e4}) oscillate with $n$  
as $n$ increases. In a system with a $g$ factor $g_s=5$, as shown in Fig. \ref{fig2}(b),
Eq. (\ref{e2}) is almost as accurate as for a system with $g_s=0$ due to the fact that the Hamiltonian Eq. (\ref{ham}) can be solved exactly when $\beta=0$. The error distribution range introduced by using Eq. (\ref{e4}) for  $g_s=5$ shrinks compared to that for $g_s=0$.
For $\alpha\ll\beta$ the approximate result is as  precise as that for  $\beta\ll\alpha$.

The density of states (DOS) is defined by
$D(E) =\sum_{nk_x\sigma} \delta(E -E_n^{\sigma})$.
Assuming a Gaussian broadening of constant width $\Gamma$ and  zero shift we obtain
\begin{equation}
D(E) =\frac{S_0}{(2\pi)^{3/2}l_c^2\Gamma} \sum_{n,\sigma}
e^{-(E - E_n^{\sigma})^{2}/2\Gamma ^{2}}.
\label{dos}
\end{equation}
In Fig. \ref{fig3}(a) the DOS is plotted as function of the energy for $\alpha=\alpha_0$, $\beta=0$, and $g_s=0$ at a magnetic field $B=0.2$ T. As discussed in our previous work \cite{wang}, it shows a beating pattern as a result of two branches of energy levels, i.e., the $+$ and $-$ ones, with different energy separations between the levels due to the spin splitting. The $m$th node
of the beating pattern is located near the $n$th Landau level when $E^-_{n+m}\simeq (E^+_{n-1}+E^+_n)/2$, which corresponds to an energy $\tilde{E}_m\simeq (2m+1)^2\hbar\omega_c/(4\sqrt{2}k_\alpha l_c)^2$. For the parameters used in Fig. \ref{fig3}, the node energies are $0.92, 2.55, 5, 8.26$, and $12.3$ meV for the first five nodes. 
For   $\beta \ll \alpha$, the spin splitting between the $+$ and $-$ branches
is enhanced and the node energy is reduced by a
factor  $1-\beta^2/\alpha^2$ as shown in Fig. \ref{fig3}(b).
In Fig. \ref{fig3}(c) and (d), the DOS  is shown as a function of the energy using the same parameters as in Fig. \ref{fig3}(a) and (b),
respectively, but with a nonzero Zeeman term  $g_s=5$.
The Zeeman term does shift the levels and the DOS along the energy axis but does not affect  the beating pattern substantially in the SOI regime discussed here; its effect on transport should be more visible in a weaker SOI regime. \cite{shen}

%Fig.3
\begin{figure}[tpb]
\vspace{-2cm}
\includegraphics*[width=140mm, height=120mm]{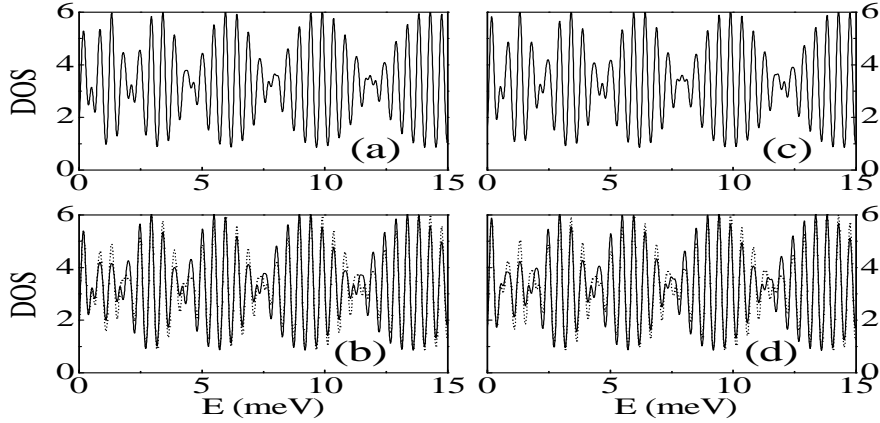}
\vspace{-3cm}
\caption{DOS as a function of energy for $\alpha=\alpha_0$ with
$\beta=0$ in (a) and (c) and $\beta=0.2\alpha_0$ in (b) and (d). The left panels are
for $g_s=0$ and the right ones for $g_s=5$. 
In panels  (b) and (d) the exact results are shown by the dotted curves and the approximate ones by the solid curves.} 
\label{fig3}
\end{figure}

For $\alpha=\beta$  the energy spectrum is a series of shifted Landau levels with degenerate
spinors as described by Eq. (\ref{e3}). The DOS appears as a series of broadened peaks with frequency $\hbar\omega_c$. When $\beta$ deviates   little from $\alpha$, however, the degenerate spinors in each Landau subband split.  As shown in Fig. \ref{fig4} and described by Eq. (\ref{e4}), the spin splitting oscillates as a function of the subband energy or Landau index and leads to a new kind of beating pattern.  

Using the asymptotic expression of the Laguerre polynomials  we obtain
\begin{eqnarray}
\label{eta}
%%%
\lambda_n\equiv\eta_n/\hbar\omega_c &\sim& (2n)^{1/4}b\sqrt{2k_sl_c}
\sin(4\sqrt{2n}k_sl_c-\pi/4)/(\sqrt{\pi}a)\\ \nonumber
&-&g_s\mu_Bm^\ast \cos(4\sqrt{n}k_sl_c-\pi/4)/(e\hbar\sqrt{2\pi\sqrt{n}k_sl_c}).
\end{eqnarray}
When the magnetic field is weak and many Landau levels are occupied, as assumed here,
the second term of  
Eq. (\ref{eta}) can be safely neglected. Then the zeros of $\eta_n$
occur for $4\sqrt{2n}k_sl_c=(p+1/4)\pi$  with $p$  integer.
The spin splitting evaluated from the exact energy spectrum (zigzagged curve) and from Eq. (\ref{eta}) (smooth curve) are plotted in Figs. \ref{fig4}(a) and (d) for $g_s=0$ and $g_s=5$, respectively.
In Fig. \ref{fig4}(a), the zeros of $\eta_n$ from Eq. (\ref{eta}) occur at $E=0.252(p+1/4)^2$ meV with
$p=0,1,2,\cdots$.
These zeros remain almost intact when the Zeeman term is considered as shown in Fig. \ref{fig4}(d). In Figs. \ref{fig4}(b) and (e)  we show the exact DOS for $g_s=0$ and $g_s=5$ respectively. We find little dependence of the beating pattern on the Zeeman term.
In Figs. \ref{fig4}(c) and (f) we plot again the DOS for $g_s=0$ and $g_s=5$, respectively,
calculated from the approximate energy spectrum Eq. (\ref{e4}) and find a good correspondence with the exact beating pattern.   The form of  the beating pattern in the DOS is determined by the ratio $\lambda_n$ of the spin splitting $\eta_n$ to the Landau energy $\hbar\omega_c$. For $\lambda_n<0.5$, as shown in Fig. \ref{fig4}, each maximum in the DOS   corresponds to a minimum in $\lambda_n=0$ and
a minimum in the DOS to a maximum in $\lambda_n$. At $\lambda_n=0.5$ a node  appears  in the beating pattern while for a $\lambda_n$ maximum larger than $0.5$ there corresponds an extra maximum in 
the DOS oscillation as shown in Fig. \ref{fig5}. Notice that the peaks of a wrap for $\lambda_n < 0.5$ appear at the centers of the Landau levels
${\cal E}^\sigma_n=(n+1/2)\hbar\omega_c-2\hbar^2 k_s^2/m^\ast$ while the peaks of a wrap for
$\lambda_n > 0.5$ appear in the middle of two adjacent Landau levels
${\cal E}^\sigma_n+1/2\hbar\omega_c$. This DOS-peak-position transition between $\lambda_n < 0.5$ wraps
and $\lambda_n > 0.5$ wraps leads to the even-odd filling factor transition in the SdH oscillation
described in the next section.
%%%

%Fig.4
\begin{figure}[tpb]
\vspace{-4cm}
\includegraphics*[width=140mm, height=150mm]{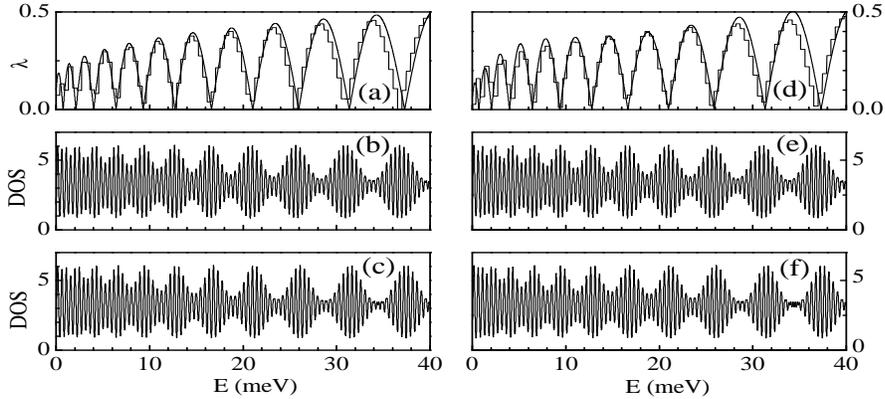}
\vspace{-3.5cm}
\caption{SOI splitting over the Landau energy $\lambda_n$ ((a) and (d)),
exact DOS ((b) and (e)), and approximate DOS 
((c) and (f)) as functions of the energy $E$.
The zigzagged curves in (a) and (d) are the exact results and
the smooth curves   the approximate one. The parameters are $\alpha=2.2\alpha_0$, $\beta=1.8\alpha_0$, and $B=0.2$T. The left panels are
for $g_s=0$ and the right ones for $g_s=5$.} 
\label{fig4}
\end{figure}

%Fig.5
\begin{figure}[tpb]
\vspace{-6cm}
\includegraphics*[width=140mm,  height=150mm]{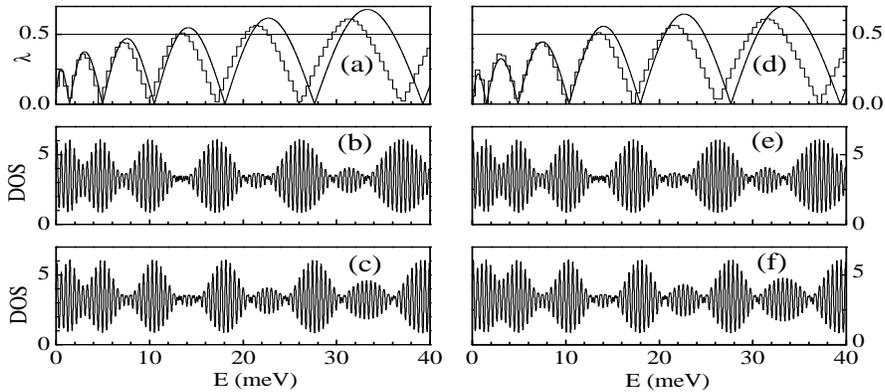}
\vspace{-3.5cm}
\caption{The same as in Fig. \ref{fig4} but for $\alpha=1.2\alpha_0$ and $\beta=0.8\alpha_0$.} 
\label{fig5}
\end{figure}
\section{Transport coefficients}
For weak electric fields $E_{\nu }$, i.e., for linear responses, and weak
scattering potentials the expression for the direct current (dc) conductivity
tensor $\sigma_{\mu \nu }$, in the one-electron approximation, reviewed in
Ref. \cite{vas2}, reads $\sigma _{\mu \nu }=$ $\sigma _{\mu \nu }^{d}+\sigma _{\mu \nu
}^{nd}$ with $\mu ,\nu =x,y,z$. The terms $\sigma _{\mu \nu }^{d}$ and $\sigma _{\mu
\nu }^{nd}$\ stem from the diagonal and nondiagonal part of the density
operator $\hat{\rho }$, respectively,
 in a given basis and $\langle J_{\mu }\rangle =Tr(\hat{\rho }J_{\mu
})=\sigma _{\mu \nu }E_{\nu }$. In general, we have $\sigma _{\mu \nu }^{d}=$
$\sigma_{\mu \nu }^{dif}+\sigma _{\mu \nu }^{col}$. The term $\sigma_{\mu \nu
}^{dif}$ describes the diffusive motion of electrons
and the term $\sigma _{\mu \nu }^{col}$ collision
contributions or hopping. The former  is given by
\begin{equation}
\sigma _{\mu\nu }^{dif}=\frac{\beta e^{2}}{S_{0}}\sum_{\zeta }
f(E_{n}^{\sigma} )[1-f(E_{n}^{\sigma} )]
\tau ^{\zeta }(E_{n}^{\sigma})
v_{\mu }^{\zeta }v_{\nu }^{\zeta },  \label{diff}
\end{equation}
where $\zeta \equiv (n,\sigma,k_{y})$ denotes the quantum numbers, $
v_{\mu }^{\zeta }=\langle \zeta |v_{\mu }|\zeta \rangle $ is the diagonal
element of the velocity operator $v_{\mu }$, and
$f(\varepsilon )$ the Fermi-Dirac function. Further,
$\tau ^{\zeta }(E_{n}^{\sigma})$ is
the relaxation time for elastic scattering,
$\beta=1/k_BT$, and $S_0$ is the area
of the system.
The term $\sigma _{\mu \nu }^{col}$ can be written  in the form

%%% If you don't need the integral over \epsilon', just remove it.
%%% We need it
\begin{equation}
\sigma _{xx}^{col}=\frac{\beta e^{2}}{2S_{0}}
\sum_{\zeta ,\zeta^\prime}
\int_{-\infty }^\infty
d\varepsilon 
\int_{-\infty }^\infty
d\varepsilon^\prime
\delta [\varepsilon -E_n^\sigma (k_y)]
\delta [\varepsilon^\prime-E_{n^\prime}^{\sigma^{\prime }}
(k_y^\prime)]
f(\varepsilon )[1-f(\varepsilon^\prime)]
W_{\zeta \zeta^\prime}(\varepsilon ,\varepsilon^\prime)
(x_\zeta-x_{\zeta^\prime})^{2},
\label{col}
\end{equation}
where $x_\zeta=\langle \zeta |x|\zeta \rangle$;
$W_{\zeta \zeta ^\prime}(\varepsilon ,\varepsilon ^\prime)$
is the transition rate.
For  elastic scattering by dilute impurities, of density $N_I$,
we have
\begin{equation}
W_{\zeta \zeta ^{\prime }}(\varepsilon ,\varepsilon ^{\prime })=\frac{2\pi
N_{I}}{\hbar S_{0}}\sum_{{\bf q}}|U({\bf q})|^{2}
|F_{\zeta \zeta^{\prime }}(u)|^{2}
\delta (\varepsilon -\varepsilon ^{\prime })
\delta_{k_y,k_y^{\prime }-q_y},
\label{rat}
\end{equation}
where $u=l_{c}^{2}q^{2}/2$ and $q^2=q_{x}^{2}+q_{y}^{2}$.
%***
%$U({\bf q})=(e^{2}/2\epsilon _{0}\epsilon )/\sqrt{q^2+k_0^2}$  is the
%Fourier transform of the screened impurity potential with $\epsilon$ the static dielectric %constant,
%$\epsilon_0$ the dielectric permittivity, and $k_0$ the screening wave vector.
$U({\bf q})$ is the
Fourier transform of the screened impurity potential
$U({\bf r})=(e^2/4\pi\epsilon_0\epsilon)e^{-k_0r}/r$,
 $\epsilon$ is the static dielectric constant,
$\epsilon_0$ the dielectric permittivity, 
and $k_0$ the screening wave vector. $U({\bf q})$ is given by
\begin{equation}
U({\bf q})
=\frac{e^{2}}{2\epsilon _{0}\epsilon }
\ \frac{1}{(2u/l_{c}^{2}+k_0^{2})^{1/2}}.
\label{uq}
\end{equation}
In the situation studied here the diffusion contribution given by
Eq. (\ref{diff}) vanishes because
the diagonal elements of the velocity operator
$v_{\mu}^{\zeta }$ vanish. Neglecting
Landau-level mixing, i. e.,  taking $n^{\prime }=n$,
and noting that $\sigma _{xx}^{col}=\sigma
_{yy}^{col}$, $\sum_{{\bf q}}=(S_{0}/2\pi )\int_{0}^{\infty
}qdq=(S_{0}/2\pi l_{c}^{2})\int_{0}^{\infty }du$, and
$\sum_{k_y}=(S_{0}/2\pi l_{c}^{2})$, we obtain
\begin{equation}
\sigma _{xx}^{col}=\frac{N_{I}\beta e^{2}}{2\pi \hbar l_{c}^{2}}
\sum_{n\sigma }\int_{0}^{\infty }du\int_{-\infty }^{\infty }
d\varepsilon
[\delta
(\varepsilon-E_{n}^{\sigma })]^{2}
f(\varepsilon)[1-f(\varepsilon)]\left| U\left( \sqrt{
2u/l_{c}^{2}}\right) \right| ^{2}\left| F_{nn}^{\sigma }(u)\right| ^{2}u.
\end{equation}
Here $\left| F_{nn}^{\sigma }(u)\right| ^{2}
=|\langle f_n^\sigma |
e^{i{\bf q}\cdot {\bf r}}
| f_n^\sigma\ \rangle|^{2}$
is the form factor.

The form factor for $\beta=0$ has been given in   Ref. \onlinecite{wang}; a similar form factor is obtained if $\alpha=0$ with the substitutions of the variables given in Sec. II A.
For $\beta\ll \alpha$ or $\alpha \ll \beta$, the form factor is obtained in a straightforward way
with the help of the above results and Eq. (\ref{e2}) or its $\alpha \ll \beta$ counterpart.
For $\alpha \approx \beta$ the form factor reads
\begin{equation}
\left| F_{nn}^\sigma(u)\right| ^{2} =e^{-u}[L_n(u)]^{2}\cos (\sqrt{2}q_x k_sl^2_c).
\label{fact}
\end{equation}

The exponential $e^{-u}$ favors small values of $u$.  Assuming
$k_0^{2}l_{c}^{2}/2\gg u$,  we may neglect the term
$2u/l_{c}^{2}$ in the denominator of Eq. (\ref{uq}) and obtain

\begin{equation}
\sigma _{yy}^{col}=\frac{N_{I}\beta e^{2}}{4\pi \hbar b}\left[ \frac{e^{2}}{
2\epsilon \epsilon _{0}}\right] ^{2}
\sum_{n\sigma }(2n+1)\int_{-\infty }^{\infty
%%% are you sure about the exponent 2 in the \delta function?
%%% sure
}d\varepsilon[\delta (\varepsilon-E_{n}^{\sigma })]^{2}
f(\varepsilon)[1-f(\varepsilon)].
\label{largeb}
\end{equation}

The impurity density $N_{I}$  determines the Landau Level broadening
$\Gamma =W_{\zeta \zeta ^{\prime }}(\varepsilon, \varepsilon^{\prime})/\hbar$.
Evaluating $W_{\zeta \zeta ^{\prime }}(\varepsilon ,\varepsilon^{\prime})/\hbar$ 
in the  $u\to 0$ limit without taking into account the SOI, we obtain
$N_{I}\approx 4\pi [(2\epsilon \epsilon _{0}/e^{2}) ]^{2}\Gamma /\hbar$.

The resistivity tensor  $\rho _{\mu \nu }$ is given in terms of the conductivity tensor.
We use the standard expressions $\rho_{xx}=\sigma _{yy}/S$ and $\rho_{yy}=\sigma_{xx}/S$,
with $S=\sigma_{xx}^2 +\sigma _{yx}^2$. In weak magnetic fields where the beating patterns
are well observed, we have $\sigma_{xy}^{nd}\approx ne/B$.

%Fig.6
\begin{figure}[tpb]
\vspace{-2cm}
\includegraphics*[width=120mm]{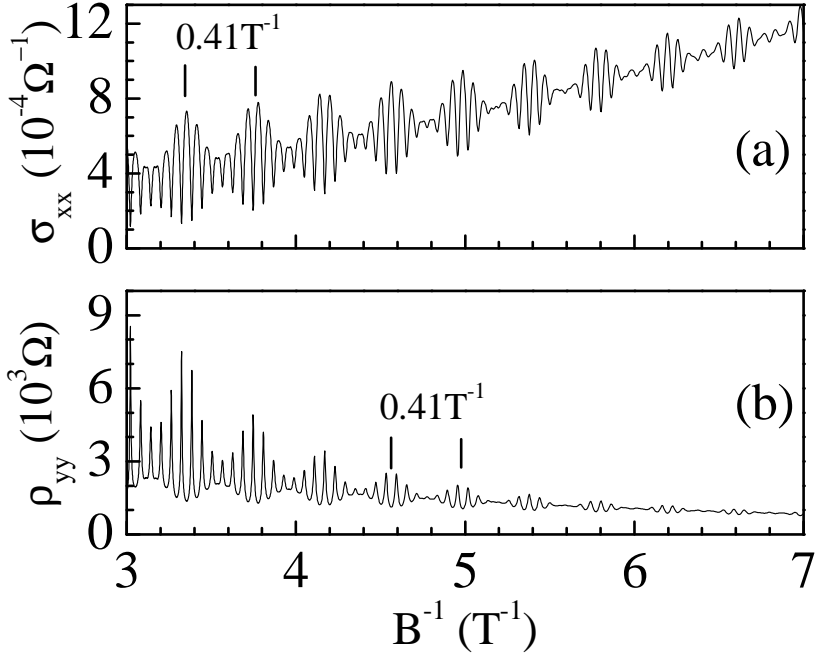}
\vspace{-3cm}
\caption{SdH oscillations of the conductivity $\sigma _{xx}$ in (a) and of the resistivity $\rho _{yy}$ in (b) vs inverse  magnetic field $B^{-1}$ for a 2DEG with $\alpha=2.2\alpha_0$, $\beta=1.8\alpha_0$,
$g_s=5$, and $n_e=8\times 10^{11}$ cm $^{-2}$.} 
\label{fig6}
\end{figure}

In weak magnetic fields a beating pattern is observed in the presence of SOI except for 
$\alpha=\beta$.
When $\beta \ll \alpha$ the beating pattern is similar to that of $\beta=0$ as described in
Ref. \onlinecite{wang} with a  period reduced by a factor $1-\beta^2/\alpha^2$.
As a function of $1/B$ the period of a beating pattern for $\alpha\approx \beta$ can be estimated from the asymptotic expression of the spin splitting as
$[4am^\ast \sqrt{2n_e/\pi}/e]^{-1}$. In Fig. \ref{fig6} we show a periodic beating pattern 
in the conductivity $\sigma _{xx}$, in (a), and   the resistivity $\rho _{yy}$,  in (b),
for a 2DEG with $\alpha=2.2\alpha_0$ and $\beta=1.8\alpha_0$.  The period of the  pattern is $B^{-1}=0.41$T$^{-1}$ and agrees well with the above estimate.

One interesting aspect of the beating pattern of the SdH oscillations is the even-odd filling factor transition \cite{edmo}, a phenomenon in which the peaks in one wrap of the beating pattern happen at even filling factors of the 2DEG while  in the next wrap occur at odd filling factors.
For $\alpha \gg\beta$ or $\alpha\ll \beta$ there is always a even-odd filling factor transition
between two consecutive beats. However, for $\alpha \approx \beta$ this transition occurs
only for $\lambda_n > 0.5$. In Fig. 7(a) we plot again the conductivity of Fig. 6(a) as a function of the filling factor $nh/eB$ and do not observe the even-odd transition. In Fig. 7(b)
the conductivity for a system with different SOI strength $\alpha=1.2\alpha_0$ and $\beta=0.8\alpha_0$ is plotted and shows the even-odd transition. 
%%% Is the following sentence obvious or more details are necessary?
%%% details are given
This difference can be explained using the DOS in Figs. \ref{fig4} and \ref{fig5}.
In Fig. \ref{fig4}, where the DOS pertaining to the parameters of Fig. \ref{fig7}(a) is shown, $\lambda_n$ is
less than 0.5 and the peaks of the DOS appear at energy $\mathcal{E}_n^\sigma$ given by
Eq. (\ref{e3}) coinciding with  the Fermi energy of the 2DEG 
for an {\it odd} filling factor. % is locates.
On the other hand, in Fig. \ref{fig5}, where the DOS of the 2DEG pertaining to the parameters of Fig. \ref{fig7}(b) is shown,
the peaks of the DOS occur at $\mathcal{E}_n^\sigma$ when $\lambda_n < 0.5$ but transit to $\mathcal{E}_n^\sigma+0.5\hbar\omega_c$, an energy corresponding to the Fermi energy
for an {\it even} filling factor and $\lambda_n > 0.5$.

%Fig.7
\begin{figure}[tpb]
\vspace{-2cm}
\includegraphics*[width=90mm]{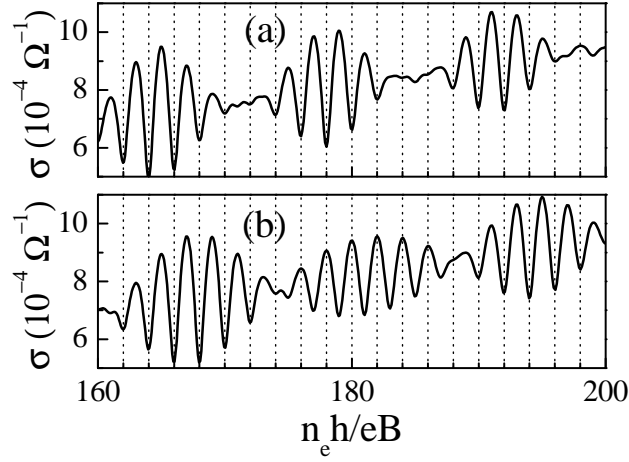}
\vspace{-2cm}
\caption{(a) The conductivity in Fig. \ref{fig6}(a) as a function of the filling factor $nh/eB$.
(b) The same as in (a) for a 2DEG with $\alpha=1.2\alpha_0$, $\beta=0.8\alpha_0$, $g_s=5$
and $n_e=8\times 10^{11}$ cm $^{-2}$.} 
\label{fig7}
\end{figure}

\section{concluding remarks}
We studied the energy band structure, DOS, and SdH oscillations of a 2DEG in the presence of a perpendicular magnetic field and of both terms of the SOI,
the Rashba (RSOI) and Dresselhaus (DSOI) terms, respectively, of strength 
$\alpha$ and $ \beta$. Besides the exact solution of the Hamiltonian
with only one of these terms present ($\alpha=0$ or $ \beta=0)$, we found an exact solution for
equal strengths  $\alpha=\beta$ provided the Zeeman coupling is negligible.
Then the band structure is just a series of
spin-degenerate Landau levels shifted by the constant $2\hbar^2k^2_s/m^\ast$, cf. Eq. (13). 
When a difference exists between the strengths $\alpha$ and $ \beta$, we obtained an approximate analytical expression
for the energy spectrum and used it to study the magnetotransport of the 2DEG.

A beating pattern of the DOS and the SdH oscillations develops but with  various characteristics
when the ratio $\alpha/\beta$  is varied. For $\alpha\gg\beta$  or $\alpha\ll\beta$,
the band structure consists of  two branches of unequally spaced levels  and
the corresponding beating patterns of the DOS and  SdH oscillations show an even-odd filling
factor transition.  
On the other hand, for $\alpha\approx\beta$  the band structure is approximately
a series of shifted Landau subbands with the spin splitting varying with
subband index. As a result, the beating patterns of the DOS and of the SdH oscillations
do not show any even-odd filling factor transition if the spin-splitting is less than
half of the Landau energy.

\begin{acknowledgments}
 This work was supported by the  Canadian NSERC Grant No.
OGP0121756.
\end{acknowledgments}


\begin{references}
\bibitem{wink}R. Winkler, {\it Spin-orbit coupling effects in two-dimensional electron and hole
systems}, Springer Tracts in Modern Phys. Vol. 191 (2003).

\bibitem{bych}  Y. A. Bychkov and E. I. Rashba, J. Phys. C {\bf 17}, 6039 (1984).

\bibitem{dres} G. Dresselhaus, Phys. Rev. {\bf 100}, 580 (1955).

%***
\bibitem{dyak} M. I. D'yakonov and V. Yu. Kachorovskii, Fiz. Tekh. Poluprovodn. {\bf 20}, 178 (1986)
[Sov. Phys. Semicond. {\bf 20}, 110 (1986).

\bibitem{verv} L. Vervoort, R. Ferreira, and P. Voisin, Phys. Rev. B {\bf 56}, R12744 (1997).

\bibitem{ross} U. Rossler and J. Kainz, Solid State Communications {\bf 121}, 313 (2002).

\bibitem{wang2}X. F. Wang, P. Vasilopoulos, and F. M. Peeters, Phys. Rev. B {\bf 65}, 165217 (2002).
\bibitem{luo}  J. Luo, H. Munekata and F. F. Fang, P. J. Stiles, Phys. Rev. {\bf 41}, 7685 (1990).

\bibitem{koga}T. Koga, J. Nitta, T. Akazaki, and H. Takayanagi, Phys. Rev. Lett. {\bf 89}, 46801 (2002).

\bibitem{gani} S. D. Ganichev, V. V. Bel'kov, L. E. Golub, E. L. Ivchenko, P. Schneider, S. Giglberger, J. Eroms, J. De Boeck, G. Borghs, W. Wegscheider, D. Weiss, and W. Prettl,
Phys. Rev. Lett. {\bf 92}, 256601 (2004).

\bibitem{here} J. J. Heremans, private communication.

\bibitem{schl1}J. Schliemann, J. C. Egues, and D. Loss, Phys. Rev. Lett. {\bf 90}, 146801 (2003).

%***
\bibitem{wang}X. F. Wang and P. Vasilopoulos, Phys. Rev. B {\bf 67}, 85313 (2003); M. Langenbuch, M. Suhrke, and U. Rossler, Phys. Rev. B {\bf 69}, 125303 (2004).

\bibitem{nit2} J. Nitta, Y. Lin, T. Akazaki, and T. Koga, Appl. Phys. Lett. {\bf 83}, 4565 (2003).  

\bibitem{nitt}  C. M. Hu, J. Nitta, T. Akazaki, H. Takayanagai,
J. Osaka, P. Pfeffer and W. Zawadzki, Phys. Rev. B {\bf 60}, 7736 (1999). 

\bibitem{shen} S. Q. Shen, M. Ma, X. C. Xie, and F. C. Zhang, Phys. Rev. Lett. {\bf 92}, 256603 (2004).

\bibitem{vas2}  P. Vasilopoulos and C. M. Van Vliet, J. Math. Phys. {\bf 25}, 1391 (1984).

\bibitem{edmo}  J. Shi, F. M. Peeters, K. W. Edmonds, and B. L. Gallagher,  Phys. Rev. B {\bf 66}, 35328 (2002).
\end{references}
\end{document}